\documentstyle [epsfig,12pt] {article}

\newcommand{\hs}{\hspace*{.375in}}
\begin{document}

\begin{figure}[ht]
\hspace*{0.95in}
\mbox{\epsfig{file=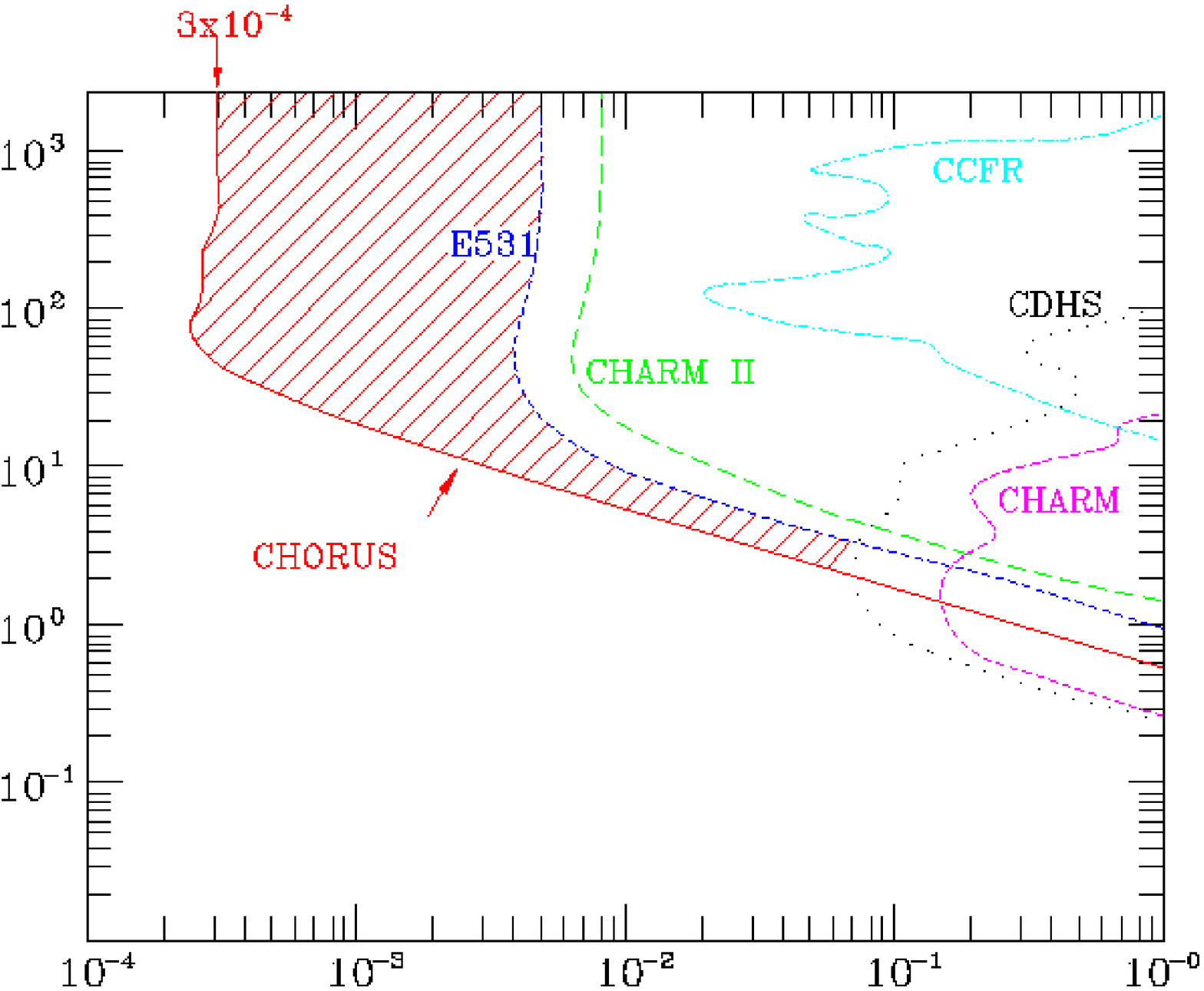,height=4.5in,width=5.0in}}\\
\caption{Previously excluded regions of the parameter space for
$\nu_\mu \rightarrow \nu_\tau$ oscillations.  Also shown is the region
where the CHORUS experiment would be able  to exclude these oscillations 
with two year's running.}
\label{explot}
\vspace*{-1in}\hspace*{3in}{\Large $\mbox{sin}^{2}\ 2 \theta$}\\
\vspace*{-3in}\\
{\Large $\Delta m^{2}$/eV$^{2}$}\\
\vspace*{3in}\\
\begin{picture}(100,100)(-210,-330)
\qbezier(50,90)(50,80)(50,70)
\qbezier(50,70)(40,60)(30,50)
\qbezier(30,50)(45,42)(60,35)
\qbezier(60,35)(30,25)(0,15)
\qbezier(0,15)(0,15)(200,-50)
\put(50,50){CCFR}
\end{picture}
\vspace*{-1in}
\end{figure}

\begin{figure}
\epsfig{figure=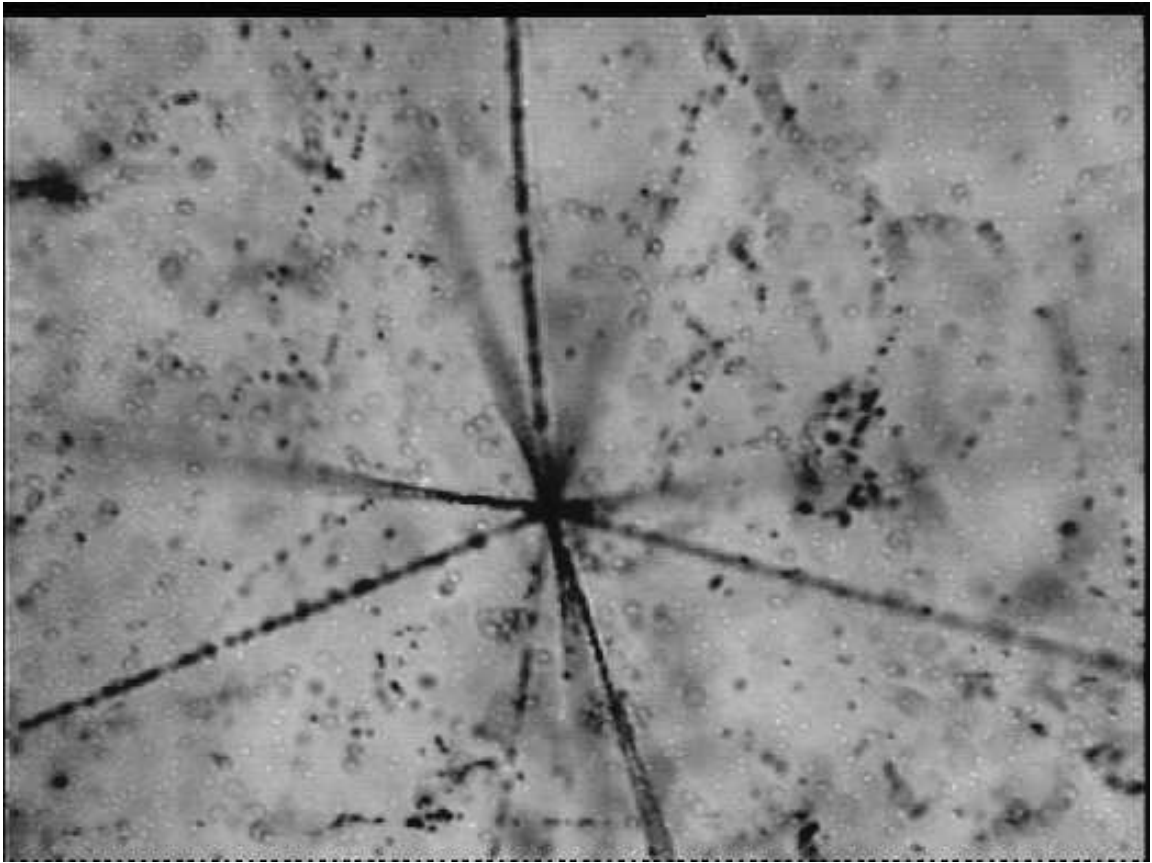,height=4.5in,width=6.0in}
\caption{An event vertex in the emulsion.  The tracks seen are due to
nuclear breakup.  Shower particles penetrate nearly
perpendicularly to the emulsion
sheet.  The image shown is approximately 100$\mu$m$\times$ 100$\mu$m.}
\label{vertex}
\end{figure}

\begin{center}
{\large {\bf STATUS OF THE SEARCH FOR $\nu_\mu\rightarrow\nu_\tau$ OSCILLATIONS
WITH THE CHORUS DETECTOR}}
\footnote{Presented at the Seventh International Workshop on
Neutrino Telescopes, Venice Italy, March 1996.  Color copies of the slides
shown may be found in:\\ 
{\tt http://choruswww.cern.ch/Publications/venice96.sheets/sheets.html}}\\
~\\
DAVID SALTZBERG\\
{\it PPE Division, CERN}\\
{\it Geneva, Switzerland}\\
E-mail: david.saltzberg@cern.ch\\
~\\
for the\\
CHORUS Collaboration\\
~\\
\end{center}
\parbox[t]{0.5in}{\hspace*{0.49in}}
\parbox[t]{5.0in}{
\begin{center}
ABSTRACT
\end{center}
~\\
The CHORUS experiment is searching for $\nu_\mu\rightarrow\nu_\tau$ 
oscillations using an 800~kg emulsion target and an electronic detector.
Two of four years of running have been completed.  The experiment will
probe small mixing angles for $\Delta m^{2}>10$~eV$^{2}$, at
least one order of magnitude better than previous limits.}\\
~\\
{\bf 1. Introduction}\\
~\\
\hs Of all the known elementary particles, perhaps the least understood are 
the neutrinos.
For example, it is not yet known if the neutrinos have a finite rest mass.
It is also not known if there is a non-zero coupling among the three different
generations of neutrinos ($e$, $\mu$ and $\tau$), such 
as exists for the quarks.
If such a coupling exists and the neutrinos have a non-zero rest mass, then
it is possible that a neutrino produced in one eigenstate of the
weak interaction ({\it e.g.}, the neutrino in $\pi\rightarrow\mu\nu_\mu$ decay)
will subsequently interact in matter as another eigenstate ({\it e.g.},
as a $\nu_e$ or $\nu_\tau$).\\
~\\
\hs The CHORUS experiment is currently searching for $\nu_\mu\rightarrow\nu_\tau$ 
oscillations.  This is a search
for the appearance of $\tau^-$ leptons produced
via $\nu_\tau$ charged-current 
interactions in a target of 800~kg of 
photographic emulsion placed in a $\nu_\mu$ beam, creating an essentially 
background-free experiment.
Two of the four years of data-taking have been completed.  The experiment
is designed to either 
discover or to exclude these oscillations with mixing,
$\mbox{sin}^{2}\ 2\theta_{\mu\tau}$, more than an order of magnitude
smaller than the previous limit
for mass differences, 
$\Delta m^2$ $\equiv$ $|m_{\nu_\tau}^2-m_{\nu_\mu}^2|>10~\mbox{eV}^{2}$.
Figure~\ref{explot}
shows the regions of this parameter space which previous 
experiments
have excluded and shows how the CHORUS collaboration
originally proposed \cite{chprop} 
it could  extend these limits
in two years of running.  CHORUS has
successfully completed two years of exposure and has extended its running
for an additional two years.~\cite{chprop2}\\
~\\
\hs It should be noted that if CHORUS sees an oscillation signal, then
it will have three discoveries at once:  direct observation
of the existence of $\nu_\tau$, violation of the conservation
of lepton generation number, and indirect proof that at least one
neutrino has a non-zero rest mass.\\
~\\
\hs This talk gives the status of the CHORUS experiment after its first
two years of data-taking.  Section~2 describes the CHORUS experiment 
and its performance.  Section~3 describes the status of the analysis of
the data
and progress towards first physics results.  Section~4 is the concluding
section.\\
~\\
{\bf 2. The experimental setup}\\
~\\
{\it 2.1 The neutrino beam}\\
~\\
\hs The CERN wide-band 
neutrino beam~\cite{wbnub} is produced by 450 GeV protons
extracted in 6-millisecond bursts
from the CERN SPS impinging on a beryllium target.  The 
secondaries, mostly pions and kaons, are focussed (positives) or 
defocussed (negatives) by magnetic toroids (``van der Meer 
horns''~\cite{wbnub}).  
The focussed particles, which
are mostly $\pi^+$, decay in a 300-meter vacuum tunnel into $\mu\nu_\mu$.
The beam is filtered by 400~meters of iron and earth.  The neutrino beam 
composition is shown in Table~\ref{nuspect}.  The contamination with directly
produced $\nu_\tau$ (from $D_S$ meson production and decay) which 
would fake an oscillation signal, yields much less than one
detected signal event over four years.\\
\begin{table}
\begin{center}
\begin{tabular}{c|c|c}
\hline\hline
type & $<E_\nu>$ & fraction\\
     &   (GeV)   &  (\%)\\
\hline
$\nu_\mu$             & 27 & 93.9\\
$\overline{\nu_\mu}$  & 22 &  5.3\\
$\nu_e$               & 48 &  0.7\\
$\overline{\nu_e}$    & 35 &  0.2\\
\hline\hline
\end{tabular}
\caption{Composition of the CERN wide-band neutrino beam used for
the CHORUS experiment.}
\label{nuspect}
\end{center}
\end{table}
~\\
{\it 2.2 $\nu_\tau$ Detection}\\
~\\
\hs CHORUS combines photographic emulsion and electronic detectors to observe
$\nu_\tau$ in the neutrino beam via its charged-current production of
a $\tau$, {\it i.e.},
\[
\nu_\tau \ + \ \mbox{nucleon} \rightarrow \tau^- \ + \ X.
\]
The principal $\tau$ decays used in the search have one charged particle,
{\it i.e.}:\\
\begin{center}
\begin{tabular}{ccll}
$\tau^-$ & $\rightarrow$ &$\mu^- \nu_\tau \overline{\nu_\mu}$ & BR=18\% \\
         &              & $h^- \nu_\tau$ X & BR=50\%,\\
\end{tabular}
\end{center}
where $h^-$ is a charged meson.\\
~\\
\hs The photographic emulsion is used as an ``active'' target since the $\tau$ 
will typically travel several hundred microns before decaying.  The resolution
of the photographic emulsion ($<1~\mu$m) is excellent for observing an
apparent break or ``kink'' in the track left by the $\tau$ candidate created
by its decay into another charged particle and neutrino(s). 
The tracking capability of the
electronic detector allows scanning of only the regions of the emulsion
where $\nu_\tau$ candidate interactions have occurred.  In addition
to the trajectories, the electronic detector measures the charges,
momenta and energies of the particles produced by the neutrino interaction
in the target and identifies muons.  This information
reduces background
and can be used to purify the sample to be scanned.
The CHORUS detector 
\begin{figure}[ht]
\epsfig{figure=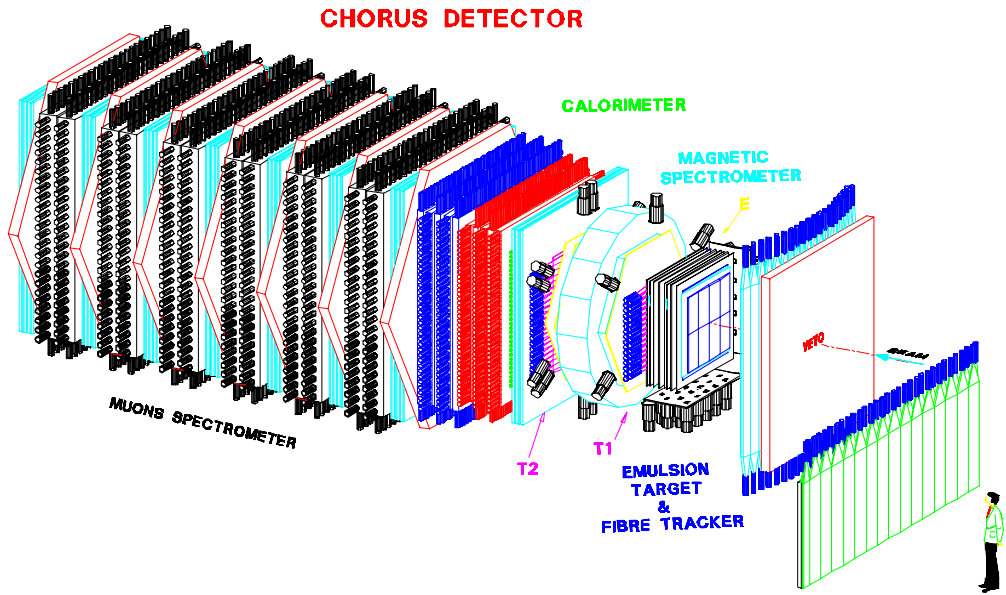,height=4.5in,width=7.0in}
\caption{Drawing of the CHORUS detector.}
\label{fulldet}
\end{figure}
is shown in Figure~\ref{fulldet}.\\
~\\
{\it 2.3 Emulsion Target}\\
~\\
\hs The target is 800~kg of photographic emulsion.  CHORUS uses
four ``stacks'' of emulsion, each consisting of 35~plates with an
area of 1.44$\times$1.44~m$^{2}$.  Each plate consists of
two 350~$\mu$m pellicles of emulsion, one on each side 
of a 90~$\mu$m acetate
backing so
that each stack is 2.8~cm deep.  Note that the particles produced
by the neutrino interaction pass roughly
perpendicularly to the emulsion planes.\\
~\\
\hs In addition to the emulsion target, 
three layers of emulsion sheets are placed
behind each ``stack'' to help connect tracks
in the electronic detector back
to tracks in the emulsion.  These tracking sheets use thin pellicles of
emulsion on both sides of a thick (800~$\mu$m) acrylic backing.  The thick
backing
provides stability and also allows a precise
angle to be determined from the hits in the emulsion on both
sides.  The first
sheet is placed directly against each emulsion stack.  The second and
third sheets are placed 40~mm and 50~mm downstream of the emulsion target
to allow the tracks to separate.\\
~\\
\hs At the end of 1994, after one year of exposure,\footnote{Each 
``year'' of running
corresponds to approximately 150 days of running between April 
and September.}
one of the four stacks
was removed and replaced by a new stack in order to start the analysis.
In 1995 all four stacks were removed and developed.  The quality of 
most of the exposed and developed emulsion looks good.  All four stacks
were replaced by new emulsions for the 1996 and 1997 runs.\\
~\\
{\it 2.4 Target Trackers}\\
~\\
\hs Planes of scintillating fiber trackers~\cite{tt} are placed
behind each set of
emulsion target and tracking sheets.  
Since these detectors are close to the
target they require good two-track resolution. Optical trackers are
employed rather than drift chambers since they do not suffer from
a left-right hit ambiguity. 
The fibers have a
500~$\mu$m diameter and are 2.2~m long. 
There are a total of 32 planes,
each consisting of seven fiber layers.  Muons crossing
a fiber produce an average of 10 scintillation photons.  These are detected
at one end of the fiber 
by image intensifiers with a photocathode
efficiency of $\sim 18\%$.  
(The other fiber end is terminated with a mirror.)
The output of the image intensifiers
is recorded by CCD cameras.  The optoelectronic readout is more
fully described in Reference~\cite{tt}.
The resolution for matching tracks in the emulsion is 
currently about 200~$\mu$m in position and 3~mrad in angle.
The two-track resolution is approximately equal to the fiber diameter.\\
~\\
{\it 2.5 Spectrometers}\\
~\\
\hs  
The experiment uses two spectrometers to determine the charge of the $\tau$ 
candidate and identify the presence of muons.
The $\nu_\tau$ charged-current interactions will
produce a negative $\tau$ giving rise to a muon or hadron also with
a negative charge.  The largest source of background ``kinks'' in
the emulsion will
be from charmed meson production
and decay; however, these mesons, when induced 
by $\nu_\mu$ will almost always have a positive charge.  
CHORUS takes advantage of the low contamination
of anti-neutrinos in the beam (which could produce negatively charged
charmed mesons) and requires ``signal'' events to
have a negatively charged $\tau$ candidate.
In addition, these charm
events will generally be 
accompanied by another muon not connected to the
``kink'' since charm is almost always made in charged-current
interactions.  Events with two muon candidates are rejected as oscillation
candidates.\\
~\\
\hs The first spectrometer
is placed upstream of the calorimeter, to be able
to determine the charge of $\tau$ ``kink'' 
candidates decaying into charged pions or low-momentum muons.
The momentum is analyzed by a hexagonally-shaped air-core magnet~\cite{hexmag}
with a field integral of $\int B\cdot\ell=1.2$~kG$\cdot$75~cm.  The
field integral is low to reduce stray fields which would adversely
affect the image intensifiers used to read out the fiber trackers.
The bend of the tracks in this field is measured with fiber tracker planes
placed in front and behind this magnet.  The resolution for 5 GeV
particles has been measured to be ${\Delta p}/{p}=16\%\oplus 4\%\ p$,
where $p$ is the momentum in~GeV.\\
~\\
\hs A muon spectrometer~\cite{muspec} is located downstream of the 
calorimeter, whose
5.2 hadronic interaction lengths filter
most particles except for muons.
The muon spectrometer consists of six iron-core toroids for a field integral
of $\int B\cdot\ell=6\times$17~kG$\cdot$50~cm.  In front or behind each magnet
there are three drift-chamber planes~\cite{drift} and eight streamer
tube planes~\cite{streamer}.
The streamer tube planes were added for the CHORUS experiment 
to improve the resolution for low-momentum muons, where multiple
scattering dominates.
The electronics for the streamer-tube planes
were upgraded with TDC readout to further
improve their position resolution.
The momentum determination is mostly limited by multiple scattering,
typically at ${\Delta p}/{p}\sim$10-15\%.\\
~\\
{\it 2.6 Calorimeter}\\
~\\
\hs Events containing a $\tau$ will have more missing momentum than 
normal $\nu_\mu$ charged-current
events since there are neutrinos produced in the $\tau$ decay.
The calorimeter~\cite{calo} is 
used to determine the energy and direction of hadron
showers.  
This information can be combined with the information from
the spectrometers to reduce the sample to be scanned in the
emulsion by selecting events with large missing momentum transverse to 
the beam direction -- enriching the sample to be scanned.\\
~\\
\hs The calorimeter consists of three sectors: electromagnetic, fine
hadronic and course hadronic with a total weight of 115 tons.
The electromagnetic sector is made of lead but 
with 20\% of its volume filled with scintillating fibers of 1~mm diameter
aligned perpendicularly to the beam direction.  Its four planes are oriented
alternately with fibers pointing horizontally and vertically.
The thickness is 21.5
radiation lengths, or 0.8 hadronic interaction lengths.
The fine hadronic calorimeter is 2.0 hadronic interaction
lengths deep and constructed similarly to the electromagnetic
sector, but with five planes and courser readout sampling.  
The course hadronic calorimeter is made of
2.4 interaction lengths of
five planes of lead-scintillator sandwich.  Finally, any
residual leakage is measured by scintillators placed inside the iron
toroids of the muon spectrometer.\\
~\\
\hs The performance of the calorimeter has been measured with pion test
beams over a large range of energies.  The resolution for electrons is
measured to be ${\sigma}/{E}=(13.8\pm0.9)\%/\sqrt{E}$.  The
resolution for pions is ${\sigma}/{E}=(32.3\pm2.4)\%/\sqrt{E}$.
The constant terms are on the order of 1\%.  The angular resolution is
approximately 60~mrad at 10~GeV.\\
~\\
The calorimeter is also instrumented with 22~streamer-tube 
planes~\cite{streamer} interspersed with the calorimeter modules.
These tracking planes
allow muons identified in the muon spectrometer to be followed back
and linked to tracks to the target region.\\
~\\
{\it 2.7 Trigger and data-taking}\\
~\\
\hs The triggering is accomplished using fast scintillator signals.  
The main trigger requires no hits upstream of the emulsion target
while hits downstream of the target must be consistent with  
tracks exiting the emulsion with an angle less than 250~mrad
with respect to the beam direction.  The timing resolution is kept better
than 2~ns to avoid vetoing events due to back-scattered particles.
Typically, the main trigger rate is 0.5 events per burst and the dead-time
is approximately 10\%.\\
~\\
\hs The data-taking 
periods of 1994 and 1995 are summarized in Table~\ref{pots}.\\
~\\
\begin{table}
\begin{center}
\begin{tabular}{c|c|c}
\hline\hline
 &1994 & 1995\\
\hline
protons on target & $0.9\times10^{19}$ & $1.1\times10^{19}$\\
\% on tape & 77\% & 88\%\\
\hline
main triggers& 400,000 & 547,000\\
charged-current events &$\sim$120,000 & $\sim$200,000\\
\hline\hline
\end{tabular}
\end{center}
\caption{Summary of the first two years of CHORUS data-taking.}
\label{pots}
\end{table}
{\bf 3. Analysis}\\
~\\
\hs One of the four ``stacks'' of emulsion was removed and replaced after
one year of running to start the analysis.  For this stack all events
are scanned for a vertex, with no kinematical preselection.   Tracks
are searched in the downstream emulsion trackers based on their predicted
positions from the target trackers.  These tracks are followed to the
emulsion sheet placed against the face of the target emulsion.  Finally
the tracks are followed into the emulsion target.
At each stage the accuracy to follow tracks improves
and the scanning area is reduced as shown in Table~\ref{scantime}.
The track in the target emulsion is followed until it stops -- presumably
at the event vertex.  
A typical vertex in the emulsion is shown in
Figure~\ref{vertex}.  
Note that the tracks seen here are nuclear fragments, the
shower particles penetrate each plate mostly perpendicularly so are not
easily visible in the figure.\\
~\\
\hs The scanning of the emulsions is currently underway
at scanning labs in Japan, Korea, Italy, Turkey and Russia using an
automatized scanning technique similar to that
described in Reference~\cite{autoscan}. 
Each 350~$\mu$m pellicle of emulsion is
scanned using a microscope and movable stage at 48 depths.  The image of the
microscope is recorded by a video camera and digitized.
Hits and tracks are found and reconstructed using the
digital signal processing technique described in Reference~\cite{autoscan}.
The current scanning efficiencies, scanning areas and scanning times are
shown in Table~\ref{scantime}.  The numbers of events found so 
far\footnote{These numbers are updates of what was presented in the
talk.} at each step are shown in Table~\ref{scannum}.\\
\begin{table}
\begin{center}
\begin{tabular}{rl|c|c|c}
\hline\hline
\multicolumn{2}{c|}{stage}& scan area & scan time & efficiency\\
     & &  ($\mu$m $\times$ $\mu$m) & (minutes) & (\%)\\
\hline
fibers&$\rightarrow$downstream emul. sheets&
  1200$\times$1200& 10 & 85\\
&$\hookrightarrow$ upstream emul. sheet &
  600$\times$600 & 1 &  80\\
&$\hookrightarrow$ target emulsion &
  120$\times$120 & 3 & 80\\
\hline\hline
\end{tabular}
\caption{Scanning area, time, and most recent efficiency to link tracks
in an event.  Scanning times are approximate.}
\label{scantime}
\end{center}
\end{table}
\begin{table}
\begin{center}
\begin{tabular}{lr}
\hline\hline
Events with predicted link into downstream emulsion~~~~& $\sim$27,000\\
Events found in downstream emulsion & $\sim$15,000\\
Events followed into upstream emulsion sheet& $\sim$10,000\\
Events followed down to vertex in target& $\sim$7,000\\
\hline\hline
\end{tabular}
\caption{Number of events at each stage of the analysis.  Currently,
all events are scanned with a kinematical selection.
The scanning
is still in progress 
so dividing the numbers at two stages does
not give the efficiency shown in Table~\protect\ref{scantime}.
Note that the FNAL-E531 result is based
on 3886 vertices before kinematical selection.}
\label{scannum}
\end{center}
\end{table}
~\\
\hs CHORUS is now in the process of digitizing the event vertices.  Also
a software algorithm is being developed to select events more likely
to have $\tau$ ``kinks''.  A software algorithm to identify ``kinks'' in
tracks is also being developed. 
Before looking systematically 
for tracks with ``kinks'' it is also necessary to correct for 
distortions of the emulsion incurred during development and drying.\\
~\\
{\bf 4. Summary}\\
~\\
\hs The CHORUS experiment has successfully taken 
two years of exposure
in a $\nu_\mu$ beam to search for $\nu_\mu\rightarrow\nu_\tau$ 
oscillations.  The experimental technique combines photographic emulsions
with a state-of-the art electronic detector.
New emulsions have been installed and two more years
of data will be taken.  The analysis of the data and search for $\nu_\tau$
interactions is underway.\\
~\\
{\bf 5. Acknowledgements}\\
~\\
\hs I thank my CHORUS colleagues for helping me prepare this talk 
and manuscript.\\
~\\
\newpage
~\\

\end{document}